# Automated SBOM-Driven Vulnerability Triage for IoT Firmware: A Lightweight Pipeline for Risk Prioritization

**Abdurrahman Tolay**
Independent Researcher, Istanbul, Türkiye
abdurrahman.tolay@outlook.com
ORCID: 0009-0000-8855-6799

January 3, 2026

### Abstract

The proliferation of Internet of Things (IoT) devices has introduced significant security challenges, primarily due to the opacity of firmware components and the complexity of supply chain dependencies. IoT firmware frequently relies on outdated, third-party libraries embedded within monolithic binary blobs, making vulnerability management difficult. While Software Bill of Materials (SBOM) standards have matured, generating actionable intelligence from raw firmware dumps remains a manual and error-prone process. This paper presents a lightweight, automated pipeline designed to extract file systems from Linux-based IoT firmware, generate a comprehensive SBOM, map identified components to known vulnerabilities, and apply a multi-factor triage scoring model. The proposed system focuses on risk prioritization by integrating signals from the Common Vulnerability Scoring System (CVSS), Exploit Prediction Scoring System (EPSS), and CISA's Known Exploited Vulnerabilities (KEV) catalog. Unlike conventional scanners that produce high volumes of uncontextualized alerts, this approach emphasizes triage by calculating a localized risk score for each finding. We describe the architecture, the normalization challenges of embedded Linux, and a scoring methodology intended to reduce alert fatigue. The study outlines a planned evaluation strategy to validate the extraction success rate and triage efficacy using a dataset of public vendor firmware, offering a reproducibility framework for future research in firmware security.

**Keywords:** IoT Security, Firmware Analysis, SBOM, Vulnerability Triage, Supply Chain Security, Embedded Linux, EPSS.

## 1 Introduction

The rapid integration of Internet of Things (IoT) devices into critical infrastructure, smart homes, and enterprise environments has expanded the attack surface available to threat actors. Unlike traditional IT assets, IoT devices are often deployed as "black box" appliances where the internal software composition is unknown to the end user. A significant portion of these devices runs on embedded Linux distributions, utilizing a complex web of open-source libraries, proprietary drivers, and legacy code.

The primary security challenge in this domain is not merely the presence of vulnerabilities, but the inability to identify and prioritize them effectively. Supply chain complexities mean that a vulnerability in a common library, such as OpenSSL or BusyBox, may propagate across thousands of device models. While the concept of a Software Bill of Materials (SBOM) has gained traction as a regulatory requirement and best practice, an SBOM is a static inventory artifact. Without context, an SBOM containing hundreds of components does not inherently improve security posture; it merely lists potential ingredients.

This paper addresses the gap between raw firmware extraction and actionable risk assessment. We propose an automated pipeline that ingests binary firmware images, extracts the file system, generates a standard SBOM, and performs vulnerability mapping. Crucially, we introduce a scoring model that moves beyond base CVSS scores by incorporating exploit maturity and exposure context to assist security analysts in triage. This paper presents the pipeline architecture and scoring method; empirical results regarding its efficacy are part of a planned evaluation.

The contributions of this paper are as follows:

- A defined methodology for automating the transition from binary firmware to structured SBOMs using open-source tooling.

- A proposed triage scoring algorithm that contextualizes vulnerability data for the IoT domain.

- A detailed discussion of the normalization challenges inherent to analyzing stripped and statically linked binaries in embedded systems.

- A reproducibility framework allowing independent researchers to replicate the analysis on public firmware datasets.

The remainder of this paper is organized as follows: Section 2 provides background on SBOMs and embedded Linux. Section 3 reviews related work. Section 4 defines the scope and threat model. Section 5 describes the dataset criteria. Section 6 through 12 detail the system architecture, extraction, generation, mapping, and scoring methodologies. Section 13 covers implementation, followed by the evaluation plan in Section 14. Sections 15 through 18 conclude the paper with discussions on limitations, ethics, and future work.

## 2 Background

### 2.1 Software Bill of Materials (SBOM)

An SBOM is a formal record containing the details and supply chain relationships of various components used in building software. Key standards include the Software Package Data Exchange (SPDX) and CycloneDX. These formats provide a machine-readable structure to describe components, licenses, and version information. In the context of IoT, generating an SBOM is complicated because the source code is rarely available to the analyst; the SBOM must be reverse-engineered from the compiled binary.

### 2.2 Embedded Linux Challenges

Embedded Linux firmware differs significantly from standard server or desktop distributions. To conserve storage and memory, developers often use:

- **BusyBox:** A single binary executable that provides several stripped-down Unix tools.

- **Stripped Binaries:** Symbol tables are removed to save space, making version identification via string matching difficult.

- **Static Linking:** Libraries are compiled directly into the executable, meaning there is no separate '.so' file to inspect for version numbers.

- **Non-Standard Filesystems:** Manufacturers use compressed read-only file systems like SquashFS, JFFS2, or UBIFS, which require specific tools to unpack.

These factors make standard package manager queries (e.g., 'dpkg' or 'rpm') impossible, necessitating filesystem-walking and signature-based heuristics for component discovery.

# 3 Related Work

The field of firmware analysis has evolved from manual reverse engineering to automated large-scale scanning.

## 3.1 Firmware Extraction and Analysis

Costin et al. conducted the first large-scale analysis of embedded firmware, identifying a correlation between shared keys and vulnerabilities across different vendors. Their work utilized 'binwalk', a standard tool for identifying and extracting embedded file system signatures. Subsequent research has focused on dynamic analysis, attempting to emulate firmware using QEMU to identify runtime vulnerabilities. However, dynamic analysis suffers from low success rates due to hardware dependencies. Our work focuses on static analysis, prioritizing coverage and speed over the depth provided by emulation.

## 3.2 SBOM and Supply Chain Security

The necessity of SBOMs has been highlighted by Executive Order 14028 in the United States, mandating greater transparency in software supply chains. Tools like 'Syft' and 'Trivy' have emerged to generate SBOMs from container images and filesystems. However, these tools are often optimized for cloud-native environments and standard package managers. Research by Feng et al. on scalable graph-based vulnerability discovery demonstrates the value of understanding component relationships, which aligns with our SBOM-driven approach.

## 3.3 Vulnerability Prioritization

Traditional vulnerability management relies heavily on the Common Vulnerability Scoring System (CVSS). However, CVSS focuses on technical severity rather than actual risk. The Exploit Prediction Scoring System (EPSS) provides a probability score indicating the likelihood of exploitation in the wild. Additionally, the CISA Known Exploited Vulnerabilities (KEV) catalog provides a binary signal of active exploitation. Our pipeline integrates these modern signals, addressing the limitations of relying solely on CVSS base scores for triage.

# 4 Scope and Threat Model

This research focuses on the **offline static analysis** of Linux-based firmware images publicly available from vendor support sites.

- **In Scope:** Firmware extraction, file system analysis, component version detection, CVE mapping, and risk scoring.
- **Out of Scope:** Dynamic analysis (emulation), hardware hacking (JTAG/UART), zero-day vulnerability discovery, and non-Linux real-time operating systems (RTOS).

The threat model assumes an attacker has access to the firmware image (either downloaded or extracted from a physical device) and seeks to identify known N-day vulnerabilities to compromise the device. The goal of the triage system is to simulate this attacker's reconnaissance phase to help defenders prioritize remediation. We define "risk" in this context as the prioritized necessity for triage, rather than a confirmed proof of exploitability.

# 5 Dataset Selection Criteria

To validate the pipeline, we define a selection strategy for a dataset comprised of firmware images from diverse device categories, including IP cameras, routers, and smart home hubs. The selection criteria prioritize: 1. **Availability:** Public download links. 2. **Architecture:** MIPS or ARM (common in IoT). 3. **OS:** Embedded Linux.

Table 1 illustrates the structure of the dataset metadata. Note that the values below are illustrative examples for the purpose of explaining the data schema.

Table 1: Illustrative Firmware Dataset Template

| Sample ID | Device Type | Vendor | Release Year | Extracted | File System |
|---|---|---|---|---|---|
| FW-SAMPLE-001 | IP Camera | Vendor A | 2023 | Yes | SquashFS |
| FW-SAMPLE-002 | SOHO Router | Vendor B | 2022 | Yes | UBIFS |
| FW-SAMPLE-003 | Smart Plug | Vendor C | 2024 | No | Encrypted |
| FW-SAMPLE-004 | NVR | Vendor A | 2021 | Yes | CramFS |

Samples that fail extraction (e.g., due to encryption or unknown compression) are logged but excluded from the downstream SBOM generation phase.

# 6 System Overview

The proposed system operates as a linear pipeline. It ingests a raw binary file and outputs a prioritized list of vulnerabilities. Figure 1 depicts the high-level architecture.

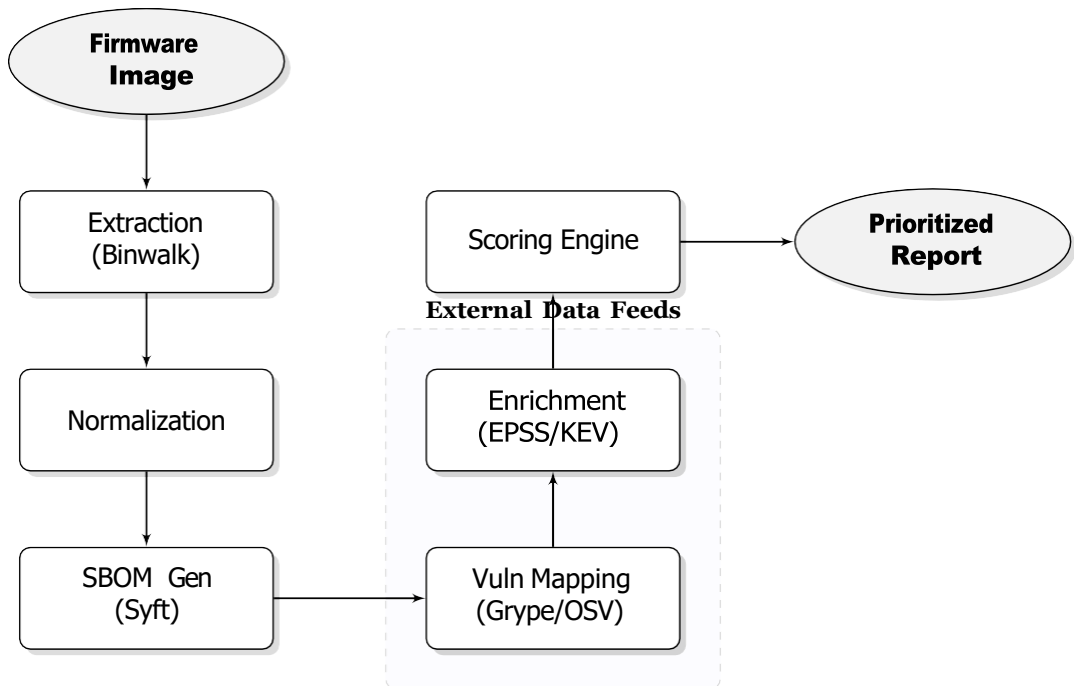

Figure 1: The Automated SBOM-Driven Triage Pipeline

The pipeline consists of four main stages: (1) Extraction, (2) SBOM Generation, (3) Vulnerability Mapping, and (4) Risk Scoring. Artifacts generated at each stage (unpacked file systems, JSON SBOMs) are preserved for auditability.

# 7 Firmware Extraction and Normalization

The first critical step is gaining access to the file system. We utilize 'binwalk', a standard firmware analysis tool, to scan the binary image for magic bytes corresponding to known file system types (e.g., SquashFS, EXT4).

## 7.1 Extraction Workflow

The extraction process is recursive. A firmware image may contain a bootloader, a kernel, and a compressed file system. Within that file system, there may be further compressed archives. The pipeline attempts to extract layers until a recognizable Linux root directory structure (containing '/bin', '/lib', '/etc') is identified.

## 7.2 Normalization

Once extracted, the data requires normalization. Symbolic links (symlinks) are a common source of infinite loops during analysis. The normalization script resolves symlinks and creates a "flat" view of the file system where possible, or flags circular dependencies to prevent the analysis engine from hanging. Additionally, permissions are normalized to ensure the analysis user has read access to all extracted artifacts.

# 8 SBOM Generation Method

With a navigable file system, the system generates the SBOM. We select the CycloneDX format (JSON specification) due to its strong support for vulnerability mapping tools and lightweight schema.

## 8.1 Component Identification

The generator scans specific directories known to house software in embedded Linux:

- **Binary Parsing:** Examining executables in '/bin' and '/usr/bin'.
- **Library Parsing:** Examining '.so' files in '/lib' and '/usr/lib'.
- **Configuration files:** Parsing specific text files (e.g., 'busybox.conf', proprietary version files).

A significant challenge is handling "unknown" versions. If a binary is stripped and lacks version strings, the system marks the version as '0.0.0-unknown'. This serves as a flag for manual review, as automated vulnerability matching against an unknown version is impossible.

# 9 Vulnerability Mapping

The SBOM is fed into a vulnerability scanner. This component compares the Package URLs (PURL) or CPE (Common Platform Enumeration) identifiers in the SBOM against databases such as the National Vulnerability Database (NVD) and GitHub Advisory Database.

## 9.1 Matching Logic and False Positives

Automated matching is prone to false positives, particularly with "backported" patches. A vendor might apply a security patch to an older version of OpenSSL without changing the version number string. To address this uncertainty, we assign a "Match Confidence" level to every finding, as detailed in Table 2.

The confidence assignment follows a strict waterfall logic. First, the scanner checks for exact metadata matches (Vendor, Product, Version) from package manager manifests (e.g., 'opkg' status files); if found, the match is **High**. If no manifest is present, the scanner inspects the binary name and available hash signatures; if the binary name is standard (e.g., 'libssl.so.1.1') but the patch level is ambiguous, the match is **Medium**. Finally, matches based purely on fuzzy filename matching (e.g., a file named 'httpd' mapping to Apache HTTP Server without corroborating evidence) are flagged as **Low**.

Table 2: Vulnerability Match Confidence Rubric

| Confidence | Criteria |
|---|---|
| **High** | Exact match of vendor, product, and version string extracted from the binary metadata or package manager manifest. |
| **Medium** | Match based on filename and hash, but version string is ambiguous or implied (e.g., generic library names). |
| **Low** | Fuzzy match based on directory structure or partial filename; high probability of false positive due to backporting. |

Findings marked as "Low" confidence are retained in the database but penalized in the final prioritization score to prevent alert fatigue.

## 10 Triage Scoring Model

A raw list of CVEs is insufficient for efficient triage. We introduce a composite Risk Priority Score (RPS) calculated for each vulnerability. The score ranges from 0 to 100.

### 10.1 Scoring Equation

The score is derived from the Base Severity ($B$), Exploit Maturity ($E$), and Exposure Context ($C$).

$$\text{RPS} = (B \times W_b) + (E \times W_e) + (C \times W_c) \tag{1}$$

Where:

- $B$: CVSS Base Score (0-10).

- $E$: Exploitability factor (0-10). Uses EPSS probability (×10) or KEV status (automatic 10 if present).

- $C$: Context factor (0-10). Defaults to 5.

- Weights: $W_b = 3.0$, $W_e = 4.0$, $W_c = 3.0$.

### 10.2 Context Factor Analysis

The Exposure Context ($C$) moves beyond static vulnerability attributes by assessing how the component is deployed in the firmware. The score increases from the default (5) to high (10) if the following signals are detected:

1. **Service Configuration:** Presence of a configuration file in '/etc' matching the component (e.g., 'lighttpd.conf').

2. **Init Scripts:** Existence of a startup script in '/etc/init.d/' or a systemd unit file enabling the service.

3. **Binary Location:** The component resides in system-critical paths like '/usr/sbin' rather than optional paths.

4. **Open Ports:** Static inference of listening ports based on configuration defaults.

## 10.3  Weight Justification and Example

The Exploitability weight ($W_e$) is highest because in the context of legacy firmware, unpatched vulnerabilities are common; the critical differentiator is whether an exploit exists.

*Illustrative Example:* Consider a vulnerability in 'dropbear' SSH (Sample ID: FW-SAMPLE-002).

- **CVSS ($B$):** 7.5 (High).

- **Exploit ($E$):** EPSS score is 0.05, but it is in the CISA KEV catalog. Therefore $E$ = 10.

- **Context ($C$):** It is a network listener (init script found). $C$ = 10.

- **Calculation:** (7.5 × 3.0) + (10 × 4.0) + (10 × 3.0) = 22.5 + 40 + 30 = 92.5.

The final RPS is 92.5, placing it in the "Critical - Immediate Action" band.

# 11  Implementation and Reproducibility

The pipeline is implemented using Python 3.9 as the orchestration layer. The primary tools integrated are:

- **Binwalk (v2.3.3):** For file system extraction.

- **Syft (v0.80+):** For generating the CycloneDX SBOM.

- **Grype/OSV-Scanner:** For vulnerability mapping.

- **EPSS API:** For fetching current EPSS values at analysis time.

To ensure reproducibility, the system uses a Dockerized environment. The Python controller scripts invoke the tools via subprocess calls, capture 'stdout' and 'stderr' for logging, and parse the JSON outputs. The scoring engine is a Python module that ingests the Grype JSON and the EPSS CSV export, merges the dataframes using Pandas, and applies the scoring logic defined in Section 10.

# 12  Evaluation Plan

As this paper presents the design and initial prototyping of the pipeline, a full-scale evaluation is planned as the next phase of research. The evaluation will assess the system based on the following metrics:

1. **Extraction Success Rate:** The percentage of firmware images from the dataset that yield a readable file system. We anticipate failures due to proprietary encryption.

2. **Component Visibility:** Comparison of the number of components identified by our pipeline versus a simple string search.

3. **Triage Efficiency:** We will measure how the RPS model changes the volume of items placed in the highest-priority band relative to CVSS-only ranking.

   4. **Manual Verification:** A random spot check of 5% of the findings to estimate the false positive rate.

# 13 Discussion

The transition from manual firmware auditing to automated SBOM-driven triage represents a necessary shift in handling the scale of IoT insecurity.

## 13.1 Practical Value

For Product Security Incident Response Teams (PSIRTs), this pipeline offers a mechanism to quickly assess the impact of a new "headline" vulnerability (e.g., Log4Shell) across a fleet of legacy devices. By archiving the SBOMs generated by this system, a PSIRT can query the database rather than re-downloading and re-scanning gigabytes of firmware images.

## 13.2 Operational Trade-offs

A purely static approach inevitably misses runtime context. A component might be vulnerable but never loaded into memory, or its vulnerable function might never be called. While our "Context" factor in the scoring model attempts to approximate this by identifying network listeners, it is not a replacement for dynamic analysis. However, the trade-off is justified by speed; this pipeline can process a firmware image in minutes, whereas dynamic emulation can take hours or days to configure per device.

## 13.3 SDLC Integration

While this paper focuses on analyzing third-party firmware, the pipeline is equally applicable within a Secure Development Lifecycle (SDLC). Developers can integrate this triage step into CI/CD pipelines to catch high-RPS vulnerabilities before the firmware is signed and released.

# 14 Limitations

The proposed system relies heavily on the quality of public vulnerability databases. If a specific version of a proprietary driver has a vulnerability that is not indexed in NVD or OSV, the system will miss it. Furthermore, the extraction capability of 'binwalk' is limited by the availability of decryption keys. If a vendor encrypts their firmware update payload, static analysis is rendered ineffective without key recovery, which often requires physical hardware attacks. Finally, the "unknown version" problem in stripped binaries remains a significant hurdle that heuristic matching can only partially solve.

# 15 Ethics Statement

This research involves the analysis of publicly available firmware updates provided by vendors for consumer download. No active scanning of live networks or devices was performed. No proprietary intellectual property was reverse-engineered beyond the standard extraction required for security interoperability analysis. All vulnerability data discussed is based on matching public CVE records; no new zero-day vulnerabilities are disclosed in this paper.

# 16  Conclusion

This paper presented a lightweight, automated pipeline for deriving actionable security intelligence from IoT firmware. By combining file system extraction, SBOM generation, and a multi-factor scoring model, we demonstrated a method to prioritize vulnerabilities based on risk rather than just technical severity. Future work will focus on integrating a "binary similarity" engine to improve version detection in stripped binaries and expanding the dataset to include Real-Time Operating Systems (RTOS) such as FreeRTOS and Zephyr.